 \documentclass[
twocolumn,showpacs, amsmath, amssymb, floatfix,
nofootinbib,wrapfloat byrevtex]{revtex4}

 \usepackage{amsmath,mathrsfs,graphicx,epstopdf,placeins,float}
 \usepackage{booktabs}
 \usepackage{multirow}

 \usepackage{pst-eps,auto-pst-pdf}

 \newcommand{\beq}[1]{\begin{equation}\label{#1}}
 \newcommand{\eeq}{\end{equation}}
 \newcommand{\bea}[1]{\begin{eqnarray}\label{#1}}
 \newcommand{\eea}{\end{eqnarray}}

\begin{document}

 \title{Schwinger pairs production in a soft-wall model}
 \author{Feng Qu}
 \email{S201406059@emails.bjut.edu.cn}
 \author{Ding-fang Zeng}
 \email{dfzeng@bjut.edu.cn}
 \affiliation{Theoretical Physics Division, College of Applied Sciences, Beijing University of Technology}
 \begin{abstract}
The Schwinger pairs production rate is calculated numerically in the soft-wall model with the help of a simpler method in determining the soft-wall's position beyond which probe strings connecting the Schwinger pairs do not fall into. The critical behaviour of the production rate and linear part in the middle region are both studied carefully. The latter manifests interesting new features. The results are compared with those in previous hard-wall models.
 \end{abstract}
 \maketitle

\section{Introduction}

Quantum field theory (QFT) tells us that the vacuum is non-empty but full of virtual particles. The Schwinger effect \cite{S1951} refers to the phenomenon that virtual particle pairs' becoming real under strong external field backgrounds. Although it has not been observed directly in laboratories due to its high requirement of the field strength, the production rate of particle pairs (Schwinger pairs) in this effect could be one important observable in the near future.

One of the motivations to study the Schwinger effect is its similarity to the Hawking radiation, i.e., charged particle pairs' production by electric field versus general particle pairs' production by gravitational field \cite{BM2010hw,RX2011hw,K2015hw}. It is intuitively worthwhile to compare the production rate of these two different phenomena. Another motivation comes from the analogue in condensed matter physics. It is known that the Schwinger effect only occurs above some critical electric field \cite{SY2013adsqcd}. This is similar to electric breakdowns in insulators which also occur above some critical field \cite{HO2013highfieldstrength} values. In this sense, we can treat the vacuum as some kind of an insulator. We expect that the study of the Schwinger pairs production rate may reveal some more information.

Since its high requirements of strong electric field imposed, the Schwinger pairs production is typical non-perturbative phenomenon. A corresponding non-perturbative tool is heavily needed if we want to calculate the production rate exactly. Gauge/gravity duality, or AdS/CFT (anti-de Sitter/conformal field theory) correspondence \cite{M1998adscft,GKP1998adscft,W1998adscft,M1999adscft,AGMOO2000adscft} is a good choice due to its functionality of translating such problems into classical gravitational ones in weakly curved space-times. In this background, Semenoff and Zarembo for the first time calculated the Schwinger pairs production rate in a quantum electrodynamics (QED) -like gauge theory using AdS/CFT \cite{SZ2011adsqcd}. They build up an $\text{AdS}_5\times\text{S}^5$ gravitational system with a probe D3-brane embedded in. The position of this probe D3-brane codes the mass of the Schwinger pair. The corresponding pairs production rate $P$ is obtained as:
\beq{}
P\sim e^{-S_{NG}-S_{B_2}}
\eeq
where $S_{NG}$ is the area of a minimal surface with a circular boundary on the probe D3-brane. $S_{B_2}$ is the contribution from some NS-NS 2-form, which represents the external field in the Schwinger effect.

The Schwinger effect  is also possible to occur in confining gauge theories such as quantum chromodynamics (QCD). Schwinger pairs production rate in a confining gauge theory is calculated using AdS/CFT by D. Kawai, Y. Sato and K. Yoshida in \cite{KSY2014adsqcd}. They use the same method as \cite{SZ2011adsqcd} but change the background into an AdS soliton one which is intended to be dual with confining gauge theories. Besides this AdS soliton background, there are many other kinds of gravitational systems that can be used as the dual of confining gauge theories. First of them is proposed by J. Polchinski and M. J. Strassler in \cite{PS2002firsthw}. They introduce an infrared cutoff on the dual $AdS_5$ background to implement confinement. Including the AdS soliton background mentioned above, such kind of modes are called hard-wall models \cite{PS2002firsthw,EKSS2005earyhw,RP2005earyhw,BT2006earyhw,SY2013adsqcd,KSY2014adsqcd,HOS2015adsqcd}.
In 2006, O. Andreev, V. I. Zakharov, A. Karch, K. Katz, D. T. Son, and M. A. Stephanov propose a new kind of model dual to confining theories in \cite{Andreev2006A,KKSS2006firstsw,AZ2006adsqcd}. They introduce a dilaton to replace the infrared cutoff in hard-wall models. O. Andreev and V. I. Zakharov find out that there is an equivalent ``wall'' in these new kind models \cite{AZ2006adsqcd}. Such new kind of models are called soft-wall models \cite{KKSS2006firstsw,GMTY2006adsqcd,AZ2006adsqcd,NTW2008adsqcd,HMY2010adsqcd}.
The existence of ``wall'' is an universal feature of the gravitational systems dual to confining gauge theories.

This paper is devoted to the calculation of Schwinger pairs production rate in the soft-wall model of AZKKSS \cite{Andreev2006A,KKSS2006firstsw,AZ2006adsqcd} using the method of \cite{SZ2011adsqcd}. It is a simple but non-trivial imitating of \cite{KSY2014adsqcd}, with the hard-wall there replaced by a soft-wall in this paper. The organization of this paper is as follows. The next section gives a simple review of calculation routines \cite{KSY2014adsqcd} in the hard-wall model with the goal of establishing necessary symbol conventions for our calculations in the soft wall model. Section \uppercase\expandafter{\romannumeral3} presents our calculation details in the soft-wall model of \cite{Andreev2006A,KKSS2006firstsw,AZ2006adsqcd}. Numerical results and comparisons with the hard-wall model will also be presented in this section. The last section is the conclusion and discussion.

\section{Review of Schwinger pairs' production rate in hard-wall models}

This section is a review of \cite{KSY2014adsqcd} in which the Schwinger pairs production rate is calculated in a hard-wall model. The gravitational system is built on the basis of an AdS soliton background composed of D3-branes \cite{HM1998adsqcd}:
\bea{hwle}
&&\hspace{-5mm}ds^2=\frac{L^2}{z^2}[(dx^0)^2+\sum^2_{i=1}(dx^i)^2+f(z)(dx^3)^2+\frac{dz^2}
{f(z)}]
\nonumber
\\
&&\hspace{1mm}+L^2ds_{S^5}^2
\\
&&\hspace{-5mm}f(z)=1-(\frac{z}{z_t})^4
\nonumber
\eea
This is written in the Euclidean signature. $L$ is the AdS radius. The singularity $z=z_t$ defines a pseudo ``horizon'' on which one spatial dimension vanishes. It is also the wall's position where the dual probe string  terminates, hence $z_t$ notation. The AdS boundary locates at $z=0$ and the inner space is $S^5$, which is not relevant in the calculation.

\begin{figure}[ht]
\begin{center}
\includegraphics[scale=0.8]{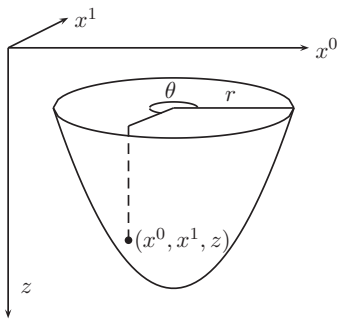}
\end{center}
\caption{~The parameterization for the cup-like minimal surface.}
\label{cup}
\end{figure}

Just as Semenoff and Zarembo's proposes \cite{SZ2011adsqcd}, the probe D3-brane locates at some intermediate position $z=z_0$ between the AdS boundary $z=0$ and the wall $z=z_t$, namely $0\leq z_0\leq z_t$, in order to describe Schwinger pairs with finite masses. The bigger $z_0$ is, the smaller mass will be, when fixing $z_t$. The minimal surface with a circular boundary (a circle on the $x^0-x^1$ plane) on the probe D3-brane can be regarded as a world-sheet of some string whose ends locate at $z=z_0$.
Since coordinates $x^0~,~x^1$ have equal rights in the line element ($\ref{hwle}$), the minimal surface should have rotational symmetry, just as a cup. Please refer to FIG.\ref{cup} and the parameterization (\ref{gauge}).
\beq{gauge}
x^0=r\cos\theta~,~x^1=r\sin\theta~,~z=z(r)
\eeq

The external field is introduced as a NS-NS 2-form. The total action of the string can be written as:
\bea{action}
&&\hspace{15mm}S_{tot}=S_{NG}+S_{B_2}
\nonumber
\\
&&\hspace{-5mm}=T_F\int d\theta\int dr\sqrt{\det\gamma}-T_F\int B_{01}dx^0\wedge dx^1
\eea
In this action formula, $S_{NG}$ is the Nambu-Goto (NG) action of the string, which is just the area of the string world-sheet; $T_F$ is the string tension; $\gamma$ is the induced metric on the string world-sheet while $S_{B_2}$ comes from the NS-NS 2-form, in which $E=T_FB_{01}$ corresponds to the external field in the Schwinger effect. The equation of motion (EOM), which describes the world-sheet configuration, can be derived by $\delta S_{tot}/\delta z(r)=0$ as follows:
\beq{hdeom}
z'+\frac{2rf(z)}{z}+rz''-\frac{r(z')^2}{2f(z)}\frac{df}{dz}(z)+\frac{(z')^3}
{f(z)}+\frac{2r(z')^2}{z}=0
\eeq

\begin{figure}[ht]
\includegraphics[scale=0.8]{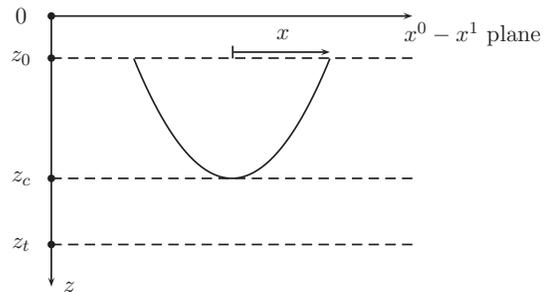}
\caption{~The radius of the minimal surface's boundary is denoted as $x$.
$z_c$ is the maximal value of $z$ on the string world-sheet. The size relationship between $z_0~,~z_{c}$ and $z_t$ is also shown here.}
\label{zc}
\end{figure}

According to \cite{SZ2011adsqcd}, the radius of this minimal surface's boundary, written as $x$~(FIG.\ref{zc}), should be fixed to such a value that the classical (on-shell) action $S_{tot}^{cl}$ extremes. This means that $x$ is not a free parameter. According to \cite{SY2013bdy}, this requirement can be replaced by a boundary condition:
\beq{addbc}
z'|_{r=x}=-\sqrt{f(z_0)(\frac{E_c^2}{E^2}-1)}
\eeq
where $E_c=T_FL^2/z_0^2$ is the critical electric field coming from potential analysis, above which the potential barrier preventing the production of Schwinger pairs vanishes \cite{SY2013adsqcd}. It implies that Schwinger pairs can be produced freely when the external field strength is bigger than $E_c$.

Referring to FIG.\ref{cup} and FIG.\ref{zc}, the remaining boundary conditions besides \eqref{addbc} can be listed as follows:
\bea{otherbc}
z|_{r=x}-z_0=&0
\nonumber
\\
z|_{r=0}-z_{c}=&0
\\
z'|_{r=0}=&0
\nonumber
\eea
Now fixing the relevant parameters the same way as \cite{KSY2014adsqcd}, $2\pi T_FL^2=10$, then solving equation (\ref{hdeom}) numerically with the above boundary conditions (\ref{addbc}) and (\ref{otherbc}), then the total action $S_{tot}^{cl}$ of (\ref{action}) can be calculated on the numerical configuration. Finally,  using ideas from reference \cite{SZ2011adsqcd}, the Schwinger pairs' production rate $P$ can be obtained as follows
\beq{hwrate}
P\sim e^{-S_{tot}^{cl}}
\eeq
All necessary numerical results in hard-wall model \cite{KSY2014adsqcd} will be reproduced in the next section in comparisons with those in the soft-wall model.

\section{Calculations in soft-wall models}

In this section we turn to the discussion of Schwinger pairs' production rate in soft-wall models. Unlike hard-wall models, positions of the ``wall'' in soft-wall models are not characterised by explicit coordinating singularity or probe field cutoff boundary conditions. So the determination of the ``wall'''s position contains technique wisdoms for the calculation of pair production rate in this model. We will provide in this section a simple method  to achieve this goal using just asymptotic analysis of the probe strings' equation of motion. This is a little different from that of reference \cite{AZ2006adsqcd}, but the conclusions are similar. As long as the the soft-wall's position is determined, the following calculations is completely parallel with those in hard-wall models. Our numerical results for the soft-wall model's calculation and comparisons with those in hard-wall models will be given in the second subsection. Some relevant and new physics explanation will also be given there.

\subsection{Schwinger pairs in soft-wall models}

Consider the following soft-wall model geometry in Euclidean signatures \cite{{Andreev2006A,AZ2006adsqcd}}:
\beq{matrix}
ds^2=\frac{L^2}{z^2}e^{\frac{cz^2}{2}}[(dx^0)^2+\sum_{i=1}^3(dx^i)^2+dz^2]
\eeq
where $L$ is the space-time radius and $c>0$ is a deformation parameter. Space-time is a pure AdS when $c\rightarrow0$. The space-time boundary locates at $z=0$.

Because there is no coordinate singularities in the background geometry of this soft wall model, the position of the wall, noted as $z=z_t$ by reference \cite{KSY2014adsqcd}, cannot be read out from ($\ref{matrix}$) directly. In the calculation of heavy quark potentials, reference \cite{AZ2006adsqcd} provide a method basing on first integrations of the probe strings' equation of motion. We provide here a more simpler one than \cite{AZ2006adsqcd} to locate the wall's position. What we need is just the EOM. Using the similar idea and parameterisations as those in hard-wall models, we derive out the EOM controlling the string world-sheet configuration in the soft-wall model as follows:
\beq{EOM}
rzz''+z(z')^3+r(2-cz^2)(z')^2+zz'+r(2-cz^2)=0
\eeq
We also use $z_{c}$ to denote the maximal value of $z$ on the string world-sheet. Please refer to FIG.\ref{zc}. Now imposing ``bottom conditions'' $z|_{r=0}=z_{c}~,~z'|_{r=0}=0$ and $z''|_{r=0}<0$~(Refer to FIG.\ref{cup} and FIG.\ref{zc}) on the EOM ($\ref{EOM}$), we will have:
\beq{}
\begin{array}{r}
\mathrm{EOM~}(\ref{EOM})
\\
z'|_{r=0}=0
\end{array}
\Big\}
\Rightarrow rzz''+r(2-cz^2)|_{r=0}=0
\eeq
\beq{}
\begin{array}{r}
\Rightarrow zz''+(2-cz^2)|_{r=0}=0
\\
z|_{r=0}>0,~z''|_{r=0}<0
\end{array}
\Big\}
\Rightarrow
\eeq
\beq{}
2-cz^2|_{r=0}>0\Rightarrow z|_{r=0}\equiv z_c<\sqrt{\frac{2}{c}}
\eeq
The above derivation tells us that $z_{c}$, the maximal value of $z$ on the string world-sheet, can't reach the value $\sqrt{2/c}$. It means that the configuration of the string whose ends locating at $z=z_0$ can't exceed the position $z=\sqrt{2/c}$. So we can identify $\sqrt{2/c}=z_t$ as the wall's position in this model. Since we divide both sides of the equation by a $0$ ($r=0$) to get the 2nd ``$\Rightarrow$'', this is not a strict proof. However, the conclusion $z_c<\sqrt{2/c}$ here is consistent with \cite{AZ2006adsqcd}. In practical numerics, when a cutoff $\epsilon$ on $r=0$ is introduced, the ``bottom condition'' would be replaced by $z|_{r=\epsilon}=z_{c}~,~z'|_{r=\epsilon}=0$ and $z''|_{r=\epsilon}<0$. In this case, the above derivations will be more acceptable.

As soon as the wall's position is located, we can perform the same procedure for numerical calculations as that in the hard-wall model. Imposing the transformation $z=z_0\zeta~,~r=z_0\rho$, we can rewrite the EOM as:
\beq{}
\rho\zeta\zeta''+\zeta(\zeta')^3+2\rho(1-\frac{z_0^2}{z_t^2}\zeta^2)(\zeta')^2
+\zeta\zeta'+2\rho(1-\frac{z_0^2}{z_t^2}\zeta^2)=0
\label{EOMhw}
\eeq
where $\zeta'=d\zeta/d\rho~,~\zeta''=d^2\zeta/d\rho^2$. The soft-wall version of boundary conditions (\ref{addbc}) and (\ref{otherbc}) are:
\bea{r}
\zeta-1|_{\rho=\frac{x}{z_0}}=0
\nonumber
\\
\zeta'+\sqrt{\frac{E_c^2}{E^2}-1}|_{\rho=\frac{x}{z_0}}=0
\label{BChw}\\
\zeta-\frac{z_{c}}{z_0}|_{\rho=0}=0
\nonumber
\\
\zeta'|_{\rho=0}=0
\nonumber
\eea
where $E_c=T_FL^2e^{z_0^2/z_t^2}/z_0^2$ is the critical electric field from \cite{STR2016esec}. Be aware that the critical electric field $E_c$ in the soft-wall model is different from that in the hard-wall model. We will choose the parameter $2\pi T_FL^2=10$ the same as that in the hard-wall model \cite{KSY2014adsqcd} and solve the boundary value problem \eqref{EOMhw}-\eqref{BChw} numerically. As long as this is done, we calculate the production rate using equation ($\ref{hwrate}$) directly.

\subsection{Numerical results and physic analysis}

\begin{figure}[ht]
\begin{center}
\includegraphics[scale=1.05]{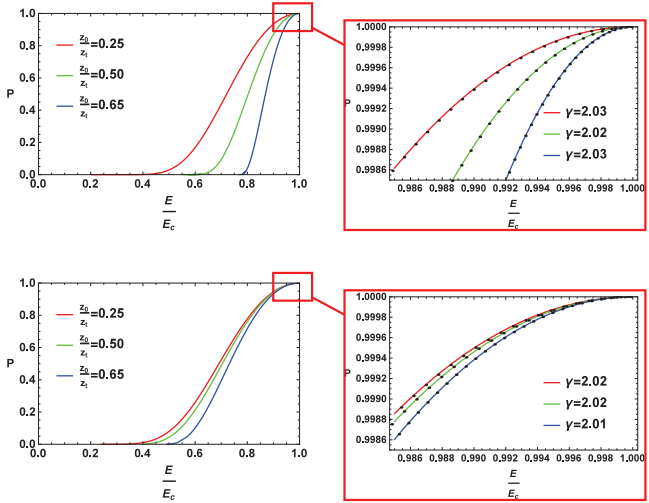}
\caption{(color online). The upper part is the production rate of Schwinger pairs in the soft-wall model. The lower part is that in the hard-wall model. Critical behavior is shown on the right side. The vertical axis is the production rate $P$. The horizontal axis is the normalized field strength $E/E_c$.}
\label{productionRateComparison}
\end{center}
\end{figure}

\begin{figure}[ht]
\begin{center}
\includegraphics[scale=0.32]{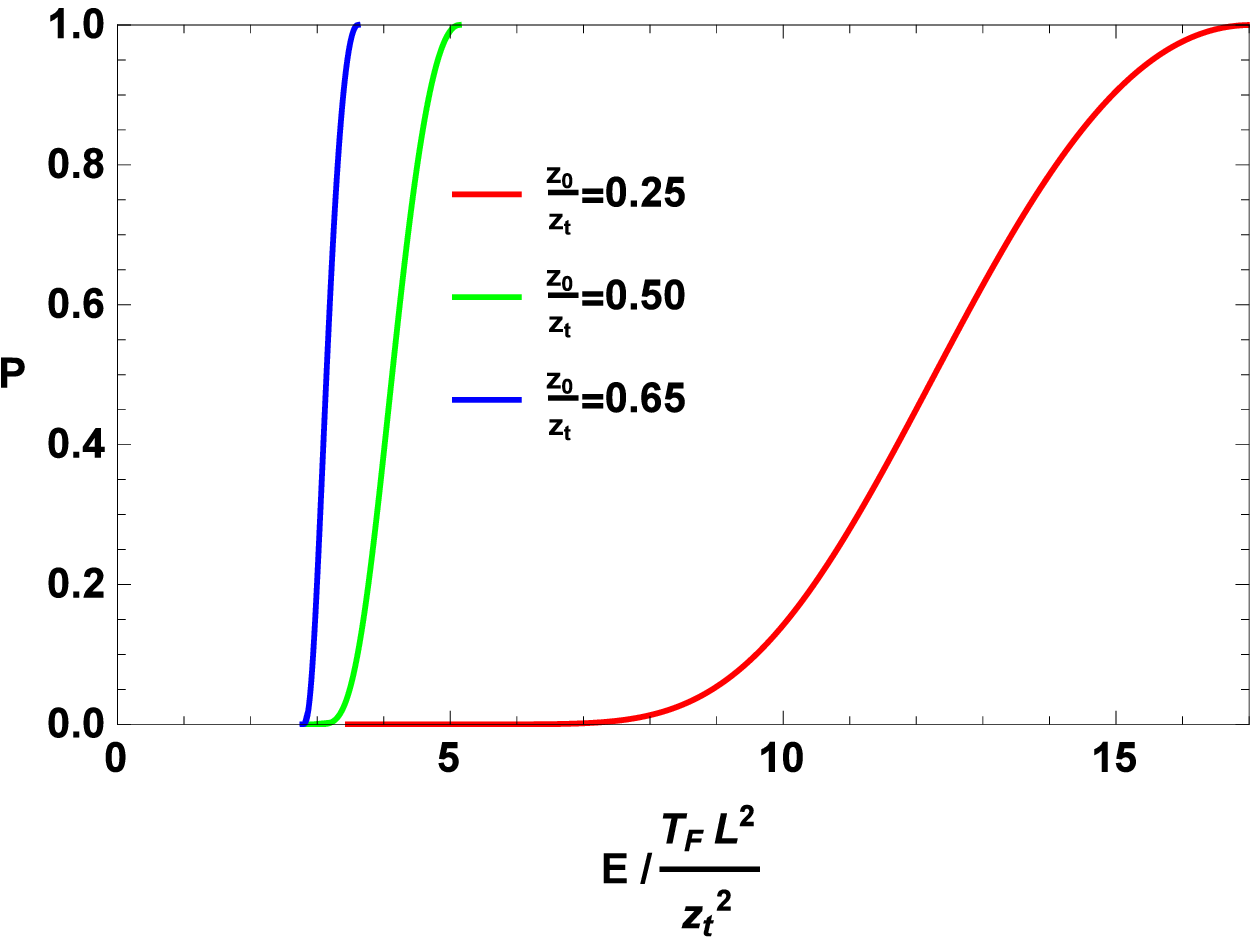}
\includegraphics[scale=0.32]{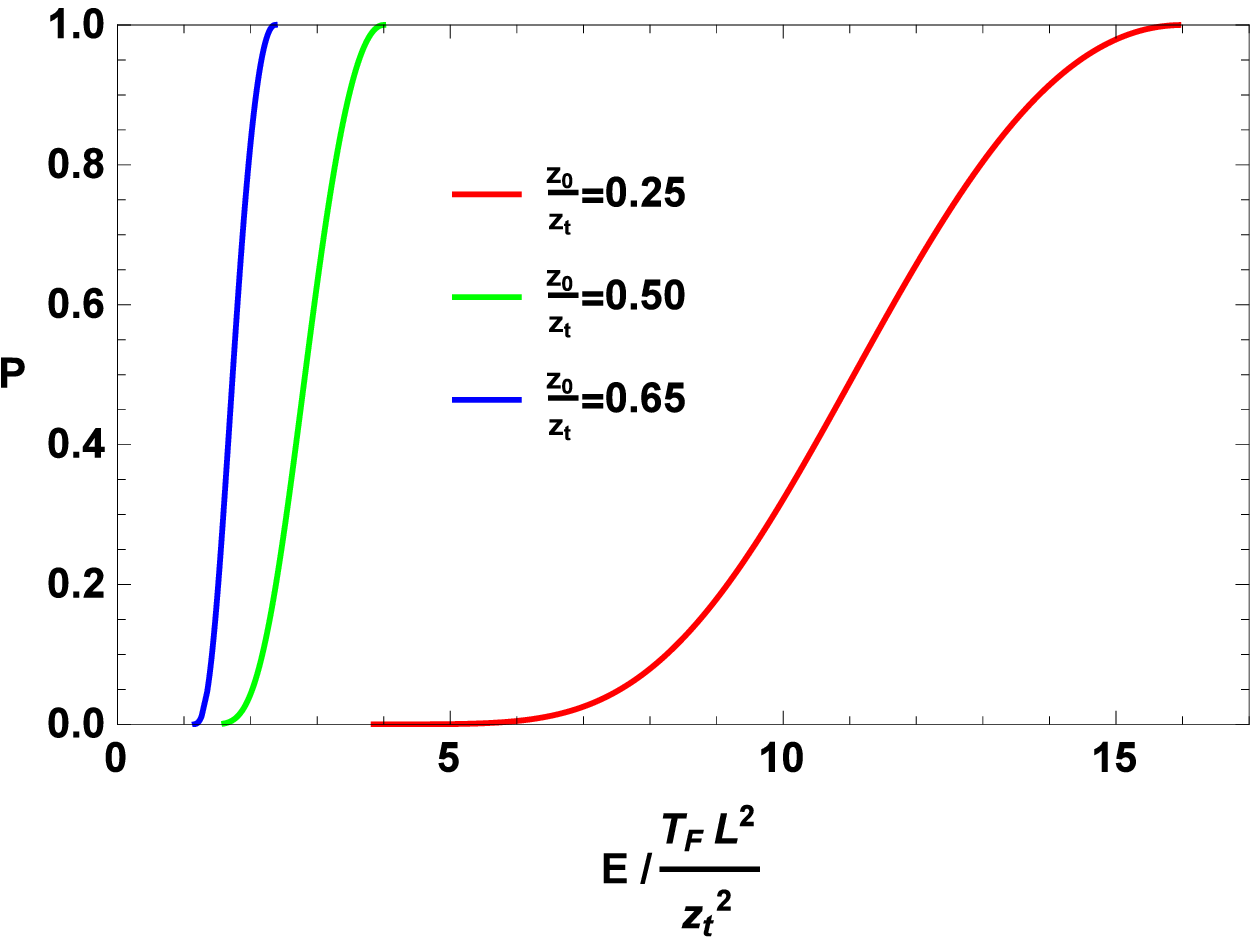}
\caption{~(color online). Production rate v.s.  abstract field strength relation. The left is for soft-wall model, while the right, hard- wall model. In both models, more smaller $z_0/z_t$ corresponds to more heavier Schwinger-pairs and more larger upper-critical field strength.}
\label{figPEabs}
\end{center}
\end{figure}

\begin{figure}[ht]
\begin{center}
\includegraphics[scale=0.32]{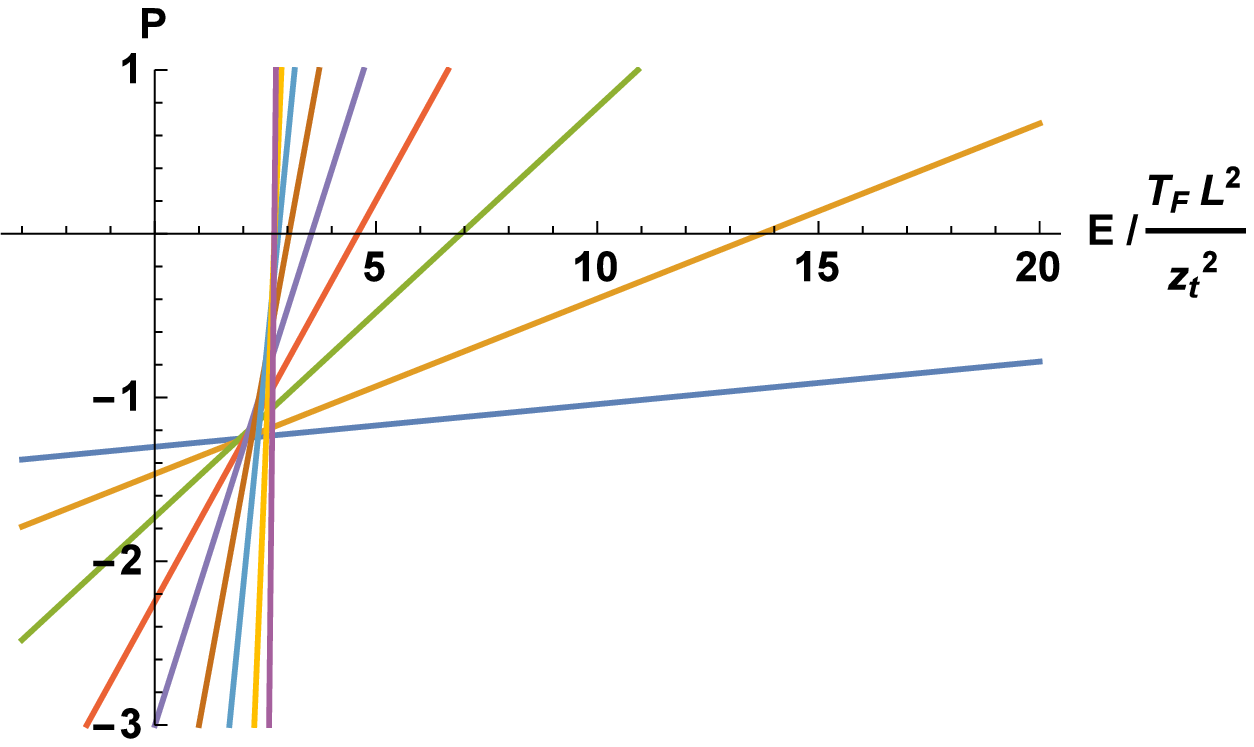}
\includegraphics[scale=0.375]{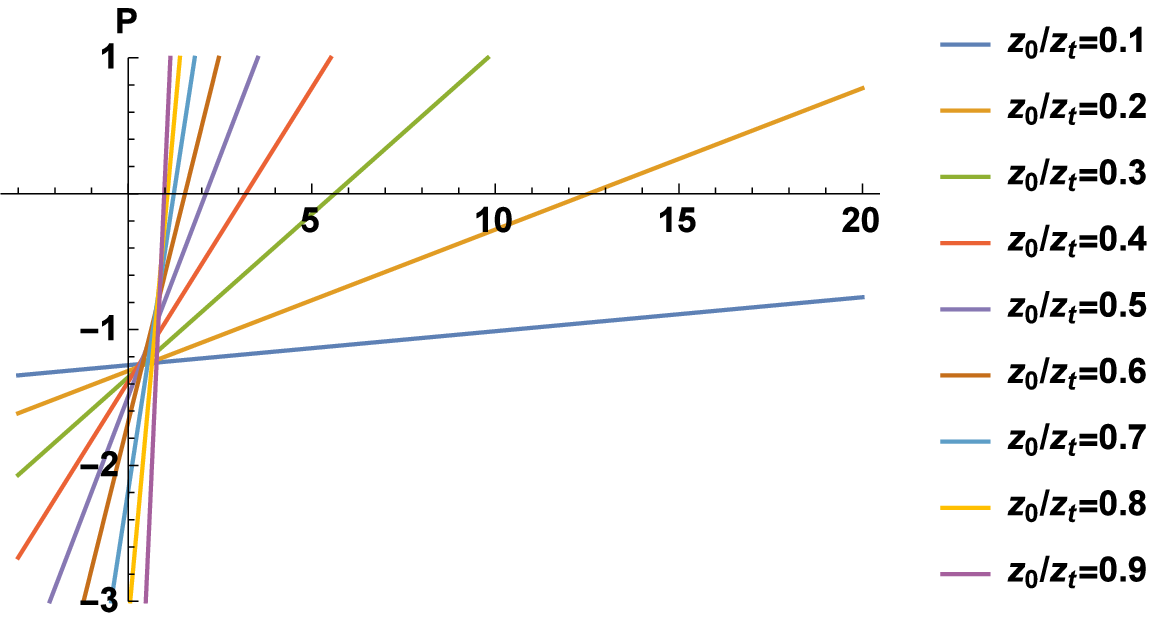}
\caption{~(color online). The linear fitting of $P$-$E$ relations as $P\in[0.2,0.8]$. The vertical axis is the production rate $P$. The horizontal axis is dimensionless field strength $E/(\frac{T_FL^2}{z_t^2})$. The left hand side is for the soft-wall model, in which all fitting lines intersect at $(E,P)=(2.5,-1.2)$. The right is for the hard-wall model, in which all fitting lines intersect at $(1,-1.2)$. }
\label{figRateFitting}
\end{center}
\end{figure}

Our numerical results for the Schwinger pairs production rate are displayed in FIG.\ref{productionRateComparison}-\ref{figRateFitting} exclusively. FIG.\ref{productionRateComparison} is mainly a comparison between the soft wall and hard wall models. From the figure, we easily see that, the $P$-$E$ line in this two models has only small quantitative but no qualitative differences. For example, both the two models display two critical values of field strength $E_c$ and $E_s$. Above $E_c$ the production rate asymptotes to $1$ while below $E_s$, the production rate vanishes approximately.

One quantitative difference between two models is, as the external field strength decreases from the upper critical value, the production rate in soft-wall models decreases more quickly than that in hard-wall models with equal parameter $z_0/z_t$. Embodying on the critical fittings
\beq{fitting}
P|_{E\rightarrow E_c}=e^{-A(1-\frac{E}{E_c})^\gamma},
\eeq
the fitting value of $A$ in the soft-wall models are (almost) always greater than that in hard-wall models when $z_0/z_t$ takes the equal values. See Table \uppercase\expandafter{\romannumeral1} for concrete numerics. However, as $z_0/z_t\rightarrow0$, the fitting value of $A$ in both models approaches the same asymptotical value $A_\mathrm{asym}\approx5$. This and the fact that $\gamma=2$ in this two models form supporting evidence for the universality conjecture of \cite{KSY2014adsqcd}.
\begin{center}
\begin{table}[!hbp]
\begin{tabular}{|c|c|c|c|c|c|c|c|}
\hline
$z_0/z_{t,\text{fixed}}$ & 0.7 & 0.6 & 0.4 & 0.2 & 0.1 & 0.03 & 0.005\\
\hline
$\gamma_{\text{soft-wall}}$ & 2.02 & 2.02 & 2.02 & 2.04 & 2.07 & 2.02 & 2.01 \\
\hline
$\gamma_{\text{hard-wall}}$ & 2.02 & 2.02 & 2.01 & 2.01 & 2.01 & 2.01 & 2.02 \\
\hline
$A_{\text{soft-wall}}$ & 35.97 & 19.62 & 9.29 & 6.64 & 6.69 & 5.61 & 5.39 \\
\hline
$A_{\text{hard-wall}}$ & 7.60 & 6.48 & 5.47 & 5.19 & 5.23 & 5.23 & 5.50 \\
\hline
\end{tabular}
\caption{Fitting parameters for the critical behavior (\ref{fitting}).}
\end{table}
\end{center}

Noting that the upper critical field strength $E_c$s in FIG.\ref{productionRateComparison} are functions of $z_0$, we cannot compare production rates of different $z_0/z_t$s at the same abstract field strength from this figure. Such comparisons are useful for our understanding physics behind the pair production but are missed in previous works. To do so, we redraw this figure with the horizontal axis $E/E_c$ replaced by $E/(\frac{T_FL^2}{z_t^2})$ in FIG.\ref{figPEabs}, where $\frac{T_FL^2}{z_t^2}$ takes equal values for all lines. By the standard dictionary of AdS/CFT, the value of $z_0/z_t$ or $z_0-z_t$ has one to one correspondence with the mass of particle members in the Schwinger pair. 
Especially, bigger $z_0/z_t$ means to smaller masses of the Schwinger pair. FIG.\ref{figPEabs} tells us that, more heavier Schwinger pairs are more difficult to be produced than the lighter ones. This is obviously consistent with our physic intuition. From FIG.\ref{figPEabs}, we can also see in addition to the upper critical $E_c$, the low critical value $E_s$ in the soft-wall model seems also larger than that in the hard-wall model. That is, Schwinger pairs are more difficult to be produced in the soft-wall model.

The $P$-$E$ relation in both FIG.\ref{productionRateComparison} and \ref{figPEabs} can also be regarded as a ``current-field strength'' relation when the field strength exceeds the lower critical value $E_s$. Physically, this is because when the Schwinger pairs are produced, external fields will drive them to form a conducting current. The bigger the production rate is, the bigger the current will be.  Examining this two figure carefully, we easily note that the linear part (P-E curves between $P=0.2$ and $P=0.8$) of this relation could be rather precisely interpreted as a sort of Ohm's laws when the vacuum conducts
\beq{}
J[\propto\,\!\!P]-J_s=\sigma(E-E_s)
\eeq
In previous works \cite{SY2013adsqcd,KSY2014adsqcd}, the lower critical field strength $E_s$' determination is a controversy question. For example, when the parameter $z_0/z_t$ takes the same value $0.25$, numeric calculation and potential analysis could cause difference as remarkable as
\beq{}
\mathrm{numeric:}~\frac{E_s}{E_c}=0.40(sw),~0.35(hw)
\eeq
\beq{}
\mathrm{potential:}~\frac{E_s}{E_c}=0.16(sw),~0.06(hw)
\eeq
However, through linear fittings of the $P$-$E$ relations in the middle part of FIG.\ref{figPEabs}, we find that almost all $P$-$E$ lines with different $z_0/z_t$ intersect on one common point. See FIG.\ref{figRateFitting} for references. The horizontal coordinate of this point seems to provide a relatively objective lower critical value of the field strength
\beq{r}
\mathrm{fitting}:~E_s=2.5(sw),1.0(hw)*\frac{T_FL^2}{z_t^2}~\mathrm{for~all}~\frac{z_0}{z_t}
\eeq
After comparing with corresponding expressions following from potential analysis \cite{SY2013adsqcd}:
\bea{r}
E_s=e*\frac{T_FL^2}{z_t^2}(sw),~
1*\frac{T_FL^2}{z_t^2}(hw)~\mathrm{for~all}~\frac{z_0}{z_t}
\eea
we can easily believe that the prediction of potential analysis should be adopted while that of direct figure reading method \cite{KSY2014adsqcd} should be given up.

\section{Conclusion and Discussion}

We calculated the Schwinger pairs production in the soft-wall model \cite{Andreev2006A,KKSS2006firstsw,AZ2006adsqcd} of confining gauge theories. Numerical results indicate that the production rate in this model is qualitatively the same but quantitatively different from that in the hard-wall models. Around the upper critical field strength $E=E_c$, the production rate in both models asymptotes to $P\sim e^{-A(1-E/E_c)^\gamma}$ with the same critical exponent $\gamma=2$. However, as the parameter $z_0/z_t\rightarrow0$, the coefficient $A$ in the soft-wall approaches the asymptotical value more quicker than that in the hard-wall model.

Relative to quantitative comparing of Schwinger pairs production rate in the soft- and hard-wall model, the more important innovation point of this work is, i) by redrawing the production rate v.s. abstract field strength i.e. $P$-$E_\mathrm{abs}$ figure, we more directly reveal that the more heavier Schwinger pairs are more difficult to be produced than the lighter ones are, see FIG.\ref{figPEabs} and captions there for references; ii) by linear fitting of middle parts of the $P$-$E$ relation in both models, we find that all the fitting lines intersect at a common point, the horizontal coordinate of which provides a rather objective determination of the lower critical field strength $E_s$ below which the pair production rate could be reasonably considered zero. This forms a very strong support for the prediction of potential analysis \cite{SY2013adsqcd}.

As discussion, the following two points could be prospected in the future. The first is, finding more definite physical interpretation for the common point of linear fitted $P-E$ lines. The second is, Schwinger pairs production in finite temperatures is also a valuable research goal. As is known, directly introducing temperature in a soliton background is difficult. However, such doing in the soft-wall model is relatively simple and directive.

\section*{Acknowledgements}

This work is supported by Beijing Municipal Natural Science Foundation, Grant. No. Z2006015201001 and partly by the Open Project Program of State Key Laboratory of Theoretical Physics, Institute of Theoretical Physics, Chinese Academy of Sciences, China.

\end{document}